\documentclass[journal]{IEEEtran}
\usepackage{cite}
\usepackage{tabularx}
\usepackage{amsmath,amssymb,amsfonts}
\usepackage{booktabs}
\usepackage{xcolor}
\usepackage{algorithm} 
\usepackage{subfigure}
\usepackage{textcomp}
\usepackage{bm}
\usepackage{soul}
\usepackage{amsmath,amsfonts}
\usepackage{algorithmic}
\usepackage{array}
\usepackage[caption=false,font=normalsize,labelfont=sf,textfont=sf]{subfig}
\usepackage{textcomp}
\usepackage{stfloats}
\usepackage{url}
\usepackage{verbatim}
\usepackage{graphicx}
\def\BibTeX{{\rm B\kern-.05em{\sc i\kern-.025em b}\kern-.08em
    T\kern-.1667em\lower.7ex\hbox{E}\kern-.125emX}}
\usepackage{balance}
\begin{document}
\title{\huge Offloading Revenue Maximization in Multi-UAV-Assisted Mobile Edge Computing for Video Stream }
\author{Bin Li and Huimin Shan

\thanks{Bin Li and Huimin Shan are with the School of Computer Science, Nanjing University of Information Science and Technology, Nanjing 210044, China (bin.li@nuist.edu.cn; 202212490251@nuist.edu.cn).}

}

\maketitle
\begin{abstract}
Traditional video transmission systems assisted by multiple Unmanned Aerial Vehicles (UAVs) are often limited by computing resources, 
making it challenging to meet the demands for efficient video processing. To solve this challenge,  
this paper presents a multi-UAV-assisted Device-to-Device (D2D) mobile edge computing system for the maximization of task offloading 
profits in video stream transmission. In particular, the system enables UAVs to collaborate with idle user devices to process video computing 
tasks by introducing D2D communications. To maximize the system efficiency, the paper jointly optimizes power allocation, 
video transcoding strategies, computing resource allocation, and UAV trajectory. The resulting non-convex optimization problem is 
formulated as a Markov decision process and solved relying on the Twin Delayed Deep Deterministic policy gradient (TD3) algorithm. 
Numerical results indicate that the proposed TD3 algorithm performs a significant advantage over other traditional algorithms 
in enhancing the overall system efficiency.
\end{abstract}

\begin{IEEEkeywords}
Mobile edge computing, video streaming, UAV, video transcoding, incentive mechanism.
\end{IEEEkeywords}

\section{Introduction}
With the demand for real-time video processing growing significantly, User Devices (UDs) have often failed to meet the real-time requirements when processing advanced video tasks.  Mobile Edge Computing (MEC) as an effective solution has been proposed by sinking computing resources to the edge network, through which UDs can offload video tasks to the edge nodes that satisfies local requests without interacting with the core network \cite{HuaTGCN2019}. As a result, network congestion is effectively alleviated, and transmission latency can be reduced. In addition, more personalized low-latency video transcoding services are provided to UDs by MEC through real-time understanding of their channel state information\cite{UAV1}.

When it comes to edge nodes, considering the high cost of construction and maintenance, as well as the limited flexibility of small BS servers, Unmanned Aerial Vehicles (UAVs) equipped with MEC servers have been deployed for their incomparable benefits, valued for their high flexibility, high-quality Line-of-Sight (LoS) channel characteristics, low cost, and more~\cite{HuaTCOM2020}. In harsh environments, computing resources can be deployed in UAVs, where UAVs act as edge computing nodes flying over ground UDs to provide timely computing services, thereby enhance computing capabilities and reduce communication latency. UAVs can be rapidly repositioned to adapt to dynamic environmental changes, making them particularly suitable for emergency situations and distant area service needs \cite{cit2}. By introducing UAVs into the MEC network, it is possible to achieve comprehensive optimization of power allocation, video transcoding, computing resource allocation, and UAV trajectory, thereby maximizing system utility~\cite{UAV2}. UAV-assisted Device-to-Device (D2D) communication can assist in resource allocation and coverage extension, improving overall system performance, while ensuring quality of service and communication continuity in dynamic environments or emergency situations~\cite{UAV3}.

Although edge computing has significant advantages for real-time tasks processing, it still faces many technical challenges. It is often costly for edge devices to perform tasks, especially since computing intensive tasks increase the energy consumption of the devices, and edge devices may be unwilling to process these computing tasks without rewards. Therefore, establishing an effective incentive mechanism to compensate for energy costs and to attract the active participation of devices is crucial. In addition, the issue of resource allocation is equally critical. Due to the limitation of bandwidth and computing resources, the processing of real-time tasks needs to be flexibly allocated between edge servers and devices. Specifically, tasks may need to be offloaded through dynamically changing links in terms of bandwidth and resources, which poses higher demands on the system's resource management. Therefore, in edge-based real-time task processing systems, the reasonable incentive mechanisms and resource allocation strategies are key to ensuring efficient system operation~\cite{incentive1}.

Existing research has mainly focused on video streaming processing in UAV-assisted MEC networks or promoted real-time video processing by devices through incentive mechanisms. However, it has not fully considered the cost issues of UAVs and other devices when collaboratively processing videos. Additionally, a naturally arising question is whether other UDs and UAVs are willing to undertake computing tasks without compensation. To fill this gap, we propose a UAV-cooperative D2D edge computing task offloading scheme for video streaming, where multiple UAVs collaboratively serve multiple UDs. This scheme leverages idle UDs and multiple UAVs to process video offloading tasks simultaneously and optimizes the video transcoding strategy of UAVs, aiming to maximize the weighted reward. 
The main technical contributions of this paper are as follows:
\begin{enumerate}
    \item We present a multi-UAV-assisted MEC network for video streaming. To attract idle UDs and UAVs to participate in video processing, an incentive mechanism is introduced. Videos are transcoded by UAVs based on a dynamic transcoding policy to reduce video processing overhead. By jointly optimizing UAV trajectory, computing resource allocation, and task offloading ratios, a system revenue maximization model is developed.
    \item To address the non-convex optimization problem with various practical constraints, it is transformed into a Markov Decision Process (MDP) model. 
    The Twin Delayed Deep Deterministic policy gradient (TD3) algorithm is proposed to find the current optimal task offloading strategy, which optimizes the weighted total utility of the system.
    \item A detailed evaluation of the TD3 algorithm is conducted and we verify its significant superiority over existing Deep Reinforcement Learning (DRL) algorithms in maximizing the total utility through numerical experiments.
\end{enumerate}

The rest of this paper is organized as follows. Section II provides related work. In Section III, the system model is presented and the optimization problem is formulated. Then, we present the MDP and TD3-based framework, as well as analyze its complexity in Section IV. Simulation results are presented in Section V. Finally, Section VI concludes the paper.

\section{Related Work}
Currently, serveral studies focus on  the task offloading problem in UAV-assisted MEC \cite{UAV4}, which has gained widespread attention due to their high mobility and easy deployment \cite{UAV5}. For example, the authors of \cite{UAV6} investigated the use of UAVs as mobile computing and caching servers for reducing task completion latency. In \cite{UAV7}, an ultra-reliable low-latency computing offloading system supporting UAVs was proposed to further reduce the latency. The authors of \cite{UAV8} proposed self-learning strategies for task offloading in UAV networks, providing new solutions for related applications. An optimization method was proposed in \cite{UAV9}, which significantly increases task offloading and completion rates in multi-UAV MEC systems by jointly optimizing three-dimensional deployment, elevation angle, computing resource allocation, and task offloading. The authors of \cite{UAV10} optimized the energy consumption and computing latency of the UAV by partially offloading tasks to the UAV in combination with D2D communication models. However, existing work primarily focused on minimizing energy consumption or latency in UAV-assisted MEC systems, which cannot be directly applied to video streaming scenarios.

With the continuous advancement of artificial intelligence, video processing is widely used in a variety of services such as motion capture and identification of personnel in disaster areas \cite{video3}. The authors of \cite{video4} studied the multi-user edge-assisted video offloading problem and analyzed it as a game theory problem to be solved by DRL. Due to the limited energy and computing resources of ground UDs \cite{video5}, it is a challenge to cope with the high computing demands required for neural network training. \cite{video6} explored a UAV-assisted video streaming system, jointly optimizing transmission power, bandwidth, and flight paths to maximize energy efficiency. In \cite{video7}, in order to minimize energy consumption, the authors proposed an energy-aware population-based mobile prediction system for UAV video transmission. However, existing research on UAV-assisted video streaming largely overlooked dynamic transcoding strategies for matching user transmission rates and transcoding bitrates, as well as the cost issues in collaborative video processing between UAVs and other UDs.

Incentive mechanisms have been studied in auction theory, game theory, and contract theory. In \cite{incentive2}, the authors utilized a swarm of UAVs to perform federated learning tasks, where the goal of the UAVs is to maximize their individual profits. In \cite{incentive3}, the authors modeled the total offloading benefit as a game process and introduced game theory to propose a reward-based distributed algorithm. Addressing parcel delivery by UAVs and ground vehicles, the authors of \cite{Incentive4} proposed a multimodal logistics framework to incentivize ground vehicles and plan UAV path. Considering the uncertainty of task offloading to UAVs, the authors of \cite{Incentive5} presented a non-cooperative game to maximize satisfaction for each user. In the context of federated learning involving UAVs and model owners, the authors of \cite{incentive6} proposed a contract matching algorithm to maximize owner profits. Nevertheless, the aforementioned studies primarily focused on profit maximization, overlooking cooperation between UAVs and UDs, as well as the capability of UDs for local processing.

\begin{figure}[t]
	\centering
	\includegraphics[width=\columnwidth]{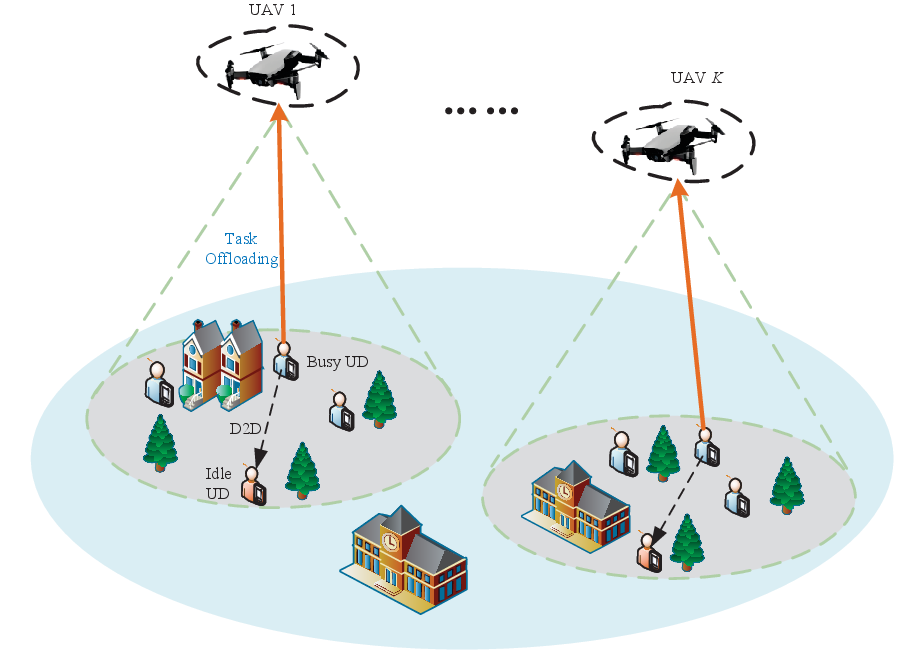}
	\caption{The system model of UAV assisted D2D edge computing network.}
	\label{fig:sys-model}
\end{figure}

Against the above research background, we propose an incentive-based UAVs collaborative D2D video stream processing system. Unlike the existing works, we consider that UAVs and idle UDs provide paid computing services for local UDs video processing. Additionally, we employ a dynamic transcoding strategy for UAVs to reduce the overall system cost.

\section{System Model and Problem Formulation}\label{s:sys}
The multi-user video transmission network is illustrated in Fig. \ref{fig:sys-model}, where multiple UAVs and UDs are introduced into the system. The UDs in the system are categorized into two types: busy UDs and idle UDs. Busy UDs in the system can choose to offload video tasks to either UAVs or idle UDs. We define the set of busy UDs as $i \in \mathcal{I} = \{1,2,\dots,I \}$, the set of idle UDs as $j \in \mathcal{J} = \{1,2,\dots,J \}$, and the set of UAVs as $k\in \mathcal{K} = \{ 1,2,\ldots,K \}$. Busy UDs send video content to UAVs and idle UDs in the form of video stream. To achieve balanced service provision and efficient utilization, the UAV flight period is defined as $T = N\partial $. In the Cartesian three-dimensional coordinate system, at time slot $n \in \mathcal{N} = \{1, 2, \ldots, N\}$, the fixed position of busy UD $i$ is given by 
$\mathbf{L}_i = \left[x_i, y_i, 0\right]^{\mathrm{T}}$,
and the fixed position of idle UD \( j \) is given by $\mathbf{L}_j = \left[x_j, y_j, 0\right]^{\mathrm{T}}$. The three-dimensional position of the UAV $k$ at time slot $n$ is $\mathbf{u}_k[n] = \left[x_k[n], y_k[n], H_k[n]\right]^{\mathrm{T}}$. Busy UDs are free to choose the target for offloading tasks within the time slot. Based on the UAVs' location information, each busy UD can select a UAV for task offloading. The factor of matching between UAVs and UDs can be expressed as follows
\begin{align}
    & \sum\limits_{k=1}^{K}{\Omega_{i,k} \le 1},\forall i \in I,\\
    &\Omega_{i,k} \in \{0,1\}, \forall i \in \mathcal{I}, k \in K,
\end{align}
where $\alpha_{i,k}=1$ if the busy UD $i$ selects to offload tasks to the UAV $k$ , and $\alpha_{i,k}=0$ otherwise.
The transition of UAV $k$ flight speed ${\bf v}_k\left[ n \right]$ and acceleration  ${\bf a}_k \left[ n \right]$ between different time slots must satisfy the following constraints
\begin{align}
    & {\bf u}_k \left[ n+1 \right] = {\bf u}_k \left[ n \right] +{\bf v}_k \left[ n \right]\partial + \frac{1}{2} {\bf a}_k \left[ n \right]\partial^2,\\
    &  \Vert {\bf a}_k \left[ n \right] \Vert = \frac{\Vert {\bf v}_k\left[ n+1\right]\Vert - \Vert {\bf v}_k\left[ n\right]\Vert}{\partial},\\
    & \Vert {\bf u}_m \left[ n \right] -{\bf u}_n \left[ n \right] \Vert^2 \ge d_{\rm {dim}}^2,m,n \in \mathcal{K},\\
    & H_k^{\min}[n] \leq H_k[n] \leq H_k^{\max}[n], \forall k \in K,
\end{align}
where ${\bf v}_k[n] \in [0, v_\text{max}]$ represents the flight speed, and ${\bf a}_k[n]$ denotes the acceleration.
\subsection{Communication Model}
To avoid interference between UDs,  orthogonal frequency division multiple access is used in D2D links. Since non-LoS transmission predominates in D2D links, the channel is modeled as a Rayleigh channel. The channel gain between the busy UD $i$ link and idle UD $j$ is defined by \cite{cit}
\begin{align}
   &{h_{i,j}}[n] = \sqrt {{\beta _0}{d_{i,j}}^{ - \chi }[n]} \left( {\sqrt {\frac{\Gamma }{{1 + \Gamma }}} \eta [n] + \sqrt {\frac{1}{{\Gamma  + 1}}} \tilde \eta [n]} \right),
\end{align}
where the channel power gain at 1 meter reference distance is represented by $\beta _0$, ${d_{i,j}} = \sqrt {||{\bf{L}}_i - {\bf{L}}_j|{|^2}}$ is the distance between busy UD $i$ and idle UD $j$, $\Gamma$ denotes the Rician factor, $\eta [n]$ and $\tilde \eta [n]$ respectively indicate the LoS channel component and the small-scale fading component. Therefore, the channel transmission rate between busy UD $i$ and idle UD $j$ is formulated as
\begin{align}
    & r_{i,j}[n] = B_0 \log_2 \left( 1 + \frac{P_{i,j}[n] \lvert{h_{i,k}}[n]\rvert^2}{\sigma_0^2} \right),
\end{align}
where $P_{i,j}[n]$ is the transmission power of busy UD $i$ when offloading to the idle UD $j$, $\sigma_0$ is the  noise power, $B_0$ is the total bandwidth of the network. Let $B_1$ be the transmission bandwidth between the UAV and the idle UDs, and ${d_{i,k}}=\sqrt{||{{\bf {u}}_k}(n)-{{\bf{L}}_i}|{|^2}} $, Thus the corresponding channel transmission rate from busy UD $i$ to UAV $k$ can be expressed as
\begin{align}
    & r_{i,k}[n] = B \log_2 \left( 1 + \frac{P_{i,k}[n] \lvert{h_{i,k}}[n]\rvert^2}{\sigma_0^2} \right),
\end{align}
where $P_{i,k}[n]$ is the transmission power of busy UD $i$ when offloading to the UAV $k$. 
\subsection{Local Computing Model}
For each busy UDs in the system, the video generated within period $T$ can be simultaneously transmitted to idle UDs and UAVs for video processing. In each time slot $n$, the size of the raw video that busy UD $i$ needs to process is denoted as $ D_i[n] $. Define $\{\varepsilon_1[n], \varepsilon_2[n], \varepsilon_3[n]\}$ as the task offloading set, where $\varepsilon_1[n]$ is the ratio of video offloaded to UAV $k$ in time slot  $n$, $\varepsilon_2[n]$ symbolizes the ratio of video offloaded to idle UD $j$, and  $\varepsilon_3[n]$ represents the ratio of video processed locally. When busy UD $i$ processes video locally, the local CPU of the busy UD performs the video parsing. The local computing resource is denoted as $f_i[n]$, and the delay to complete the video task is expressed as
\begin{align}
    t_{\text{local},i}[n] &= \frac{\varepsilon_3[n]D_i[n]C[n]}{f_i[n]}.
\end{align} 

The computing energy consumption for busy UD $i$ is defined by
\begin{align}
    E_{\text{local},i}[n] &= \kappa f_i^2[n] \varepsilon_3[n] D_i[n] C[n],
\end{align}
where $\kappa$ is the influence factor of chip architecture on CPU processing, $C[n]$ represents the CPU cycles required for local video processing.
\subsection{Offloading Computing Model}
Considering different channel conditions, in order to avoid the delay of video playback caused by the long transcoding time, the UAV $k$ selects the appropriate transcoding rate. Suppose the initial video of UDs can be transcoded from the original bitrate to levels denoted by $\{b_1, b_2, \ldots, b_O\}$. The local video has a uniform original bitrate of 2.75 Mbps and is transcoded to $b_O$ bitrate through the UAV's dynamic selection strategies.
\subsubsection{Offloading to UAVs}
When busy UD $i$ offloads video processing tasks to UAV $k$, due to the typically powerful computing capabilities of UAVs, which far exceed those of UDs, the computing task from busy UD $i$ to UAV $k$ can usually be completed entirely within one time slot. It is worth noting that since UAVs are usually battery-powered, the energy consumption of the UAV $k$ during flight for task execution also needs to be considered. The flight energy consumption of UAV $k$ in the $n^{th}$ time slot is calculated as $E_k^\text{fly}[n] = p_k^\text{fly}[n] \partial$, where $p_k^\text{fly}[n]$ represents the propulsion power of UAV $k$ during flight, modeled as
\begin{align} 
    p_k^\text{fly}[n] &= \frac{1}{2} d_c \rho g A_\mathrm{c} ||v_k[n]||^3 + P_\mathrm{a} \left(1 + \frac{3||v_k[n]||^2}{U_\text{tip}^2}\right) \notag \\
    &\quad + P_\mathrm{b} \left(\sqrt{1 + \frac{||v_k[n]||^4}{4v_\mathrm{f}^2}} - \frac{||v_k[n]||^2}{2v_\mathrm{f}^2}\right)^{\frac{1}{2}},
\end{align}
where $P_\mathrm{a}$ represents the power of UAV blades, $P_\mathrm{b}$ denotes the induced power during hover, $v_\mathrm{f}$ signifies the average rotor speed, $\rho$ stands for air density. $U_{\mathrm{tip}}$ denotes the tip speed of the blade, $d_\mathrm{c}$ indicates the fuselage drag ratio, $A_\mathrm{c}$ denotes the rotor area, and $g$ represents the rotor solidity.
The delay for offloading tasks from a busy UD $i$ to UAV $k$ is given by
\begin{align} 
    t_{\text{off},i}[n] = \frac{\varepsilon_1[n] D_i[n]}{\sum\limits_{k=1}^{K} \Omega_{i,k} r_{i,k}[n]}.
\end{align}

The energy consumption generated during the offloading process is given by
\begin{align}
    E_{\text{off},i}[n] = P_{{i,k}}[n] t_{\text{off},i}[n].
\end{align}

The CPU cycles required for UAV $k$ to process the video are given by $C_k[n] = m_1 b_O^{m_2}[n]$, where $m_1$ and $m_2$ are constants related to hardware. Considering UAV $k$ 's  requirement for video quality, according to \cite{24}, the time consumed for transcoding and subsequent computing processing of the video is given by
\begin{align}
    & t_k \left[ n \right] = \frac{{C_k}[n]}{f_k \left[ n \right]},
\end{align} 
where $f_k[n]$ denotes the computing resources required by UAV $k$ to process videos in time slot $n$. The energy consumption during transcoding is expressed as
\begin{align}
    & E_{k}[n] = s_1 f_k[n]^{y_1} t_{z,k},
\end{align} 
where the coefficients $s_1$ and $y_1$ are related to the energy consumption per unit bit associated with the UAV CPU. The size of the transcoded video is $D_k'[n]$. The delay for the UAV $k$ to process the transcoded video is formulated as
\begin{align}
    t_{k,i}[n] = \frac{ D_k'[n] C_k[n]}{\sum\limits_{k=1}^{K} \Omega_{i,k} f_k[n]}.
\end{align} 

The energy consumption generated during the offloading process is defined by
\begin{align}
    E_{k,i}[n] = \kappa f_k^2[n]D_k'[n] C_k[n].
\end{align}
\subsubsection{Offloading to idle UDs}
for the D2D execution mode, the video of the busy UD $i$ is first offloaded to the idle UD $j$ through a D2D link, and then the idle UD $j$ performs the task processing. The transmission delay for offloading to the idle UD $j$ is formulated as
\begin{align}
    t_{\text{off},ij}[n] = \frac{\varepsilon_2[n] D_i[n]}{r_{i,j}[n]}.
\end{align}

 The energy consumption during transmission is given by
 \begin{align}
    E_{\text{off},ij}[n] = P_{i,j}[n] t_{\text{off},ij}[n].
\end{align}

The delay for the idle UD $j$ to process the video is expressed as
\begin{align}
    t_{i,j}[n] = \frac{ \varepsilon_2[n] D_i[n] C[n]}{f_j[n]}.
\end{align}

The energy consumption during computing is formulated as
\begin{align}
    E_j[n] = \kappa f_j^2[n] \varepsilon_2[n] D_i[n] C[n].
\end{align}
\subsection{Resource Pricing Model}
In a multi-UAV-assisted D2D computing offloading system, UAVs and idle UDs gain utility by selling their computing resources to busy UDs. However, they must consume their own energy to process the offloaded tasks. Therefore, this paper defines the utility function for UAVs as follows
\begin{align}
&U_K = \sum_{k=1}^K \left( f_k[n] p_k[n] - \beta_k E_{k}[n] - \beta_k E_k^\text{fly}[n] - \beta_k E_{k,i}[n] \right) \notag \\
&\quad \quad = \sum_{k=1}^K \left( f_k[n] p_k[n] - \beta_k s_1 f_k[n]^{y_1} t_{z,k} - \beta_k \partial p_{\text{fly}}[n]\right. \notag \\
&\quad \quad \quad \left.  -  \beta_k \kappa f_k^2[n]D_k'[n] C_k[n]\right) ,
\end{align}
where $p_k[n]$ represents the price of computing resources provided by UAV $k$ to busy UD $i$, and $\beta_k$ represents the inconvenience factor. According to \cite{25}, $\beta_k = 1 / (1 - \varepsilon_1[n])$. The utility function also decreases with the inconvenience factor $\beta_k$, as UAVs are unwilling to share computing resources due to limited battery or scarcity of computing resources. The utility function for idle UDs is defined as the revenue obtained from selling computing resources after deducting the energy consumption costs, and is expressed as
\begin{align}
    &U_J  = \sum_{j=1}^J \left( f_j[n] p_j[n] - \beta_j E_j[n]  \right) \notag \\
&\quad \quad = \sum_{j=1}^J \left( f_j[n] p_j[n] - \beta_j \kappa f_j^2[n]  \varepsilon_2[n] D_i[n] C[n]\right),
\end{align}
where $p_j[n]$ represents the price of the computing resources provided by the idle UD $j$ to the busy UD $i$, and $\beta_j$ is the inconvenience factor of the energy consumption for the idle UDs. The busy UD $i$ can obtain utility from the computing resources purchased from UAV $k$ and the idle UD $j$, but it should pay for the computing resources. In addition, the busy UD $i$ has its own computing resources to prcess tasks. The utility function of the busy UDs is given by
\begin{align}
   & U_I = \sum_{i=1}^{I} \left( u_i f_i[n] - \beta_i \left( E_{\text{local},i}[n] - E_{\text{off},i}[n] - E_{\text{off},\text{ij}}[n] \right) \right) \notag \\
    &\quad \quad + \sum_{j=1}^{J} \left( u_{j,i} f_j[n] - f_j[n] p_j[n] \right) + \sum_{k=1}^{K} \left( u_{k,i} f_k[n]\right. \notag \\
    & \quad \quad \left. - p_k[n] f_k[n]\right),
\end{align}
where $u_i=f_i^{\text{max}} / \left[ (f_i^{\text{max}} + f_k^{\text{max}} + f_j^{\text{max}}) / 3 \right]$, $u_{k,i}=f_k^{\text{max}} / f_i^{\text{max}}$, $u_{j,i}=f_j^{\text{max}} / f_i^{\text{max}}$ are the incentive factors for the busy UD $i$, UAV $k$, and idle UD $j$ to process video tasks, respectively. $\beta_i$ is the inconvenience factor of energy consumption for the busy UDs processing local tasks. The busy UDs can gain more profit from the resource-rich UAVs and idle UDs.
\subsection{Problem Formulation}
Our goal is to maximize the utility reward of the system over the entire time period by jointly optimizing the offloading ratio ${\bm \varepsilon} \triangleq \{ \varepsilon_1[n],\varepsilon_2[n],\varepsilon_3 [ n ], \forall n \in \mathcal{N} \}$, UAVs computing resources, idle UDs computing resources, busy UDs computing resources ${\bf f}_\iota \triangleq \{ f_i[n], f_j[n], f_k[n], \forall n \in \mathcal{N}, i \in \mathcal{I}, j \in \mathcal{J}, k \in \mathcal{K} \}$, UAVs flight trajectories ${\bf u} \triangleq \{ {\bf u}_k (n), \forall n \in \mathcal{N}, k \in \mathcal{K} \}$, UAVs computing resource prices ${\bm p}_r \triangleq \{ p_k[n], \forall n \in \mathcal{N}, k \in \mathcal{K}\}$, and idle UDs computing resource prices ${\bm p}_j \triangleq \{ p_j[n], \forall n \in \mathcal{N}, j \in \mathcal{J}\}$, in order to maximize the system revenue over the entire period $T$. The system revenue is defined as the weighted sum of the UAVs, idle UDs, and busy UDs. It is expressed as follows
\begin{align}
& {Q} = {\omega _1}{U_{K}} + {\omega _2}{U_{J}} + {\omega _3}{U_I} ,
\end{align}
where $\omega_1$, $\omega_2$, and $\omega_3$ are the utility weights for the UAVs, idle UDs, and busy UDs, respectively. Therefore, the optimization problem can be formulated as
\begin{subequations}\label{P:0}
    \begin{align}
        &\mathop {\max }\limits_{{{\bf{p}}_r},{{\bf{p}}_j},\bm{\varepsilon},\bm{\omega},\bf{f}_\iota,\bf{u}} \;\sum\limits_{n = 1}^N {Q} \\
        \text{s.t.}~
        &\Vert {\bf{u}}{_p}[n] - {\bf{u}}{_q}[n]||^{2} \ge d_{\dim }^2,p,q \in \mathcal{K},\label{P0:eq_distance}\\
        &(1), (2),\label{P0:eq_mark}\\
        &{\varepsilon _1}[n] \in \left\{ {0,1} \right\},{\varepsilon _2}[n] \in \left\{ {0,1} \right\},{\varepsilon _3}[n] \in \left\{ {0,1} \right\},\label{P0:eq_rho1}\\
        &{\varepsilon _1}[n] + {\varepsilon _2}[n] + {\varepsilon _3}[n] = 1,\label{P0:eq_rho2}\\
        &{\omega _1} + {\omega _2} + {\omega _3} = 1,\label{P0:eq_weight}\\
        &{\bf{u}}[1] = {\bf{u}}[N],\label{P0:eq_u}\\
        &H_k^{\min}[n] \leq H_k[n] \leq H_k^{\max}[n], \forall k \in K,\label{P0:eq_hei}\\
        &0\le P_{i,k}\left[ n \right] \le P_{i,k}^{\text{max}},\forall n \in \mathcal{N}, i \in \mathcal{I}, k \in \mathcal{K},\label{P0:eq_puav}\\
        &0\le P_{i,j}\left[ n \right] \le P_{i,j}^{\text{max}},\forall n \in \mathcal{N}, i \in \mathcal{I}, j \in \mathcal{J},\label{P0:eq_pd}\\
        &0\le f_\iota\left[ n \right] \le f_\iota^{\text{max}},\forall n \in \mathcal{N}, \iota \in \{\mathcal{I}, \mathcal{J}, \mathcal{K}\},\label{P0:eq_fi}\\
        &p_r^{\text{min} } \le {p_r} \le p_r^{\text{max} }, \label{P0:eq_price1}\\
        &p_j^{\min } \le {p_j} \le p_j^{\text{max} }, \label{P0:eq_price2} \\
        &\sum\limits_{n = 1}^N {({E_{k}}[n] + {E_{k,i}}[n] + E_k^{\text{fly}}[n]) \le {E_r}, \forall i \in \mathcal{I}, k \in \mathcal{K},\label{P0:eq_E}} \\
        & \| \mathbf{v}_k[n] \| \leq v_{\text{max}}, \forall n \in \mathcal{N}, k \in \mathcal{K},\label{P0:eq_vk}
    \end{align}
\end{subequations}
where $\bm{\varepsilon} = \{\varepsilon_1, \varepsilon_2, \varepsilon_3\}$, $\bm{\omega}=\{\omega_1, \omega_2, \omega_3\}$. $P_{i,k}^{\text{max} }$ symbolizes the maximum transmission power for busy UD $i$ to offload to UAV $k$, $P_{i,j}^{\max }$ symbolizes the maximum transmission power for busy UD $i$ to offload to idle UD $j$.
Constraint \eqref{P0:eq_distance} denotes the minimum safety distance limitation between UAVs. 
Constraint \eqref{P0:eq_mark} indicates that a UD is restricted to connect to at most one UAV. 
Constraint \eqref{P0:eq_rho1} and constraint \eqref{P0:eq_rho2} represent the offloading ratio for busy UDs. 
Constraint \eqref{P0:eq_weight} expresses the weight factors among the UAVs, idle UDs, and busy UDs, reflecting the influence of these three parts on the total revenue. 
Constraint \eqref{P0:eq_u} restricts the start and end points of the UAV flight trajectory.
Constraint \eqref{P0:eq_hei} restricts the altitude at which UAVs can fly.
Constraint \eqref{P0:eq_puav} and constraint \eqref{P0:eq_pd} limit the transmission power of UDs. 
Constraint \eqref{P0:eq_fi} ensures that the computing resources provided by idle UDs, busy UDs, and UAVs do not exceed the device limits. 
Constraint \eqref{P0:eq_price1} and constraint \eqref{P0:eq_price2} limit the resource unit price for UAVs and idle UDs.
Constraint \eqref{P0:eq_E} ensures that the energy consumed by the UAV for flying, transcoding, and computing in all time slots does not exceed the maximum battery capacity, where $E_r$ represents the battery capacity of UAV $k$.
Constraint \eqref{P0:eq_vk} is UAVs' speed limitations.

\section{TD3-Based Cooperative Edge Computing Offloading Algorithm for UAV Video Streams}

Due to the complex decision variables and constraints involved in problem \eqref{P:0}, solving it using traditional convex optimization techniques is challenging. DRL, a fusion of deep learning and reinforcement learning techniques, aims to address complex tasks characterized by high-dimensional state and action spaces \cite{cit1}. The TD3 algorithm is an advanced deep reinforcement learning algorithm that enables an agent to make optimal decisions in complex environments by interacting with the environment, thereby maximizing long-term cumulative rewards.
Therefore, this paper proposes a multi-UAV-assisted D2D MEC network training framework that integrates the TD3 algorithm to maximize system efficiency. This framework optimizes the objective function under the premise of satisfying the constraints, overcoming the difficulties of traditional methods in high-dimensional state and action space training, and achieving real-time online decision-making and efficient resource allocation. In this network, the UDs and UAVs working in collaboration. We describe the offloading and computing process as a model-free, transition probability-free MDP. In the MDP, the agent interacts with the dynamic environment and continuously adjusts its strategy to maximize cumulative rewards. At each time step, the following are the definitions of states, actions, and rewards.

\subsection{State Space}
In the multi-UAV-assisted D2D MEC system presented in this paper, the state space is determined by the UDs, UAVs, and their environment. We define the state space as ${s_n}$, and the system state at time slot $n$ is defined as
\begin{align}
& {s_n} = \left\{ {{\mathbf{L}}_i[n],{D_i}[n],C(n)}, \mathbf{u}_k[n], {E_{kr}}[n]\right\} ,
\end{align}
where ${\mathbf{L}}_i[n]$ denotes the location of busy UD $i$ at time slot $n$, ${D_i}[n]$ symbolizes the amount of video tasks generated by busy UD $i$ at time slot $n$, ${\bf{u}}_k[n]$ symbolizes the location information of UAV $k$ at time slot $n$, $C[n]$ symbolizes the CPU resources required for local video processing, and $E_{kr}[n]$ symbolizes the remaining energy of UAV $k$ at time slot $n$.

\subsection{Action Space}
The agent needs to determine a set of actions according to the present system state and the observed environment. The action space $a_n$ can be represented as 
\begin{align}
    & a[n] = \left\{ \left[ \varepsilon_1[n], \varepsilon_2[n], \varepsilon_3[n] \right], f_i[n], f_j[n], f_k[n], p_k[n], p_j[n], \right. \notag \\
    & \quad \quad \quad\left. \left[ \omega_1, \omega_2, \omega_3 \right], {{\bf{v}}_k}[n], b \right\}, \label{eq:action_space}
\end{align}
where $\{\varepsilon_1[n], \varepsilon_2[n], \varepsilon_3[n]\}$ represent the task offloading ratios to the UAV $k$, idle UD $j$, and local processing at time slot $n$, respectively. $f_i[n]$ denotes the computing resources of busy UD $i$ at time slot $n$, $f_j[n]$ indicates the computing resources of idle UD $j$ at time slot $n$, $f_k[n]$ indicates the computing resources of UAV $k$ at time slot $n$, $p_k[n]$ and $p_j[n]$ represent the unit price of computing resources provided by UAV $k$ and idle UD $j$ to busy UD $i$, respectively. $\{\omega_1, \omega_2, \omega_3\}$ denote the weight ratios of busy UDs, idle UDs, and UAVs in the total utility gain, $\mathbf{v}_k[n]$ indicates the speed of the UAV $k$, and $b$ is the transcoding policy.
\subsection{Reward}
The objective of the TD3 architecture is to maximize the reward function by optimizing the actor and critic neural networks. To solve problem \eqref{P:0}, our reward is defined as the difference between the weighted sum of utility gains for all UDs and UAVs compared to the constraint penalty. The system reward at time slot $n$ is given by
\begin{align}
& r[n] = Q - F ,
\end{align}
where $Q$ is the revenue-weighted sum of the UAVs, idle UDs, and busy UDs at time slot $n$. $F$ represents the penalty for exceeding constraints. When constraint \eqref{P0:eq_distance} is not satisfied, meaning that the UAV exceeds the minimum safe distance, a penalty $F_1$ is set. When constraint \eqref{P0:eq_E} is not satisfied, meaning that the energy consumed by UAVs flight, transcoding, and computing in all time slots exceeds the maximum battery capacity, a penalty $F_2$ is set. When constraint \eqref{P0:eq_vk} is not satisfied, meaning that UAVs flight speed exceeds the maximum flight speed limit, a penalty $F_3$ is set. Therefore, the total penalty is $F = F_1 + F_2 + F_3$.
\begin{figure}[t]
    \centering
    \includegraphics[width=\columnwidth]{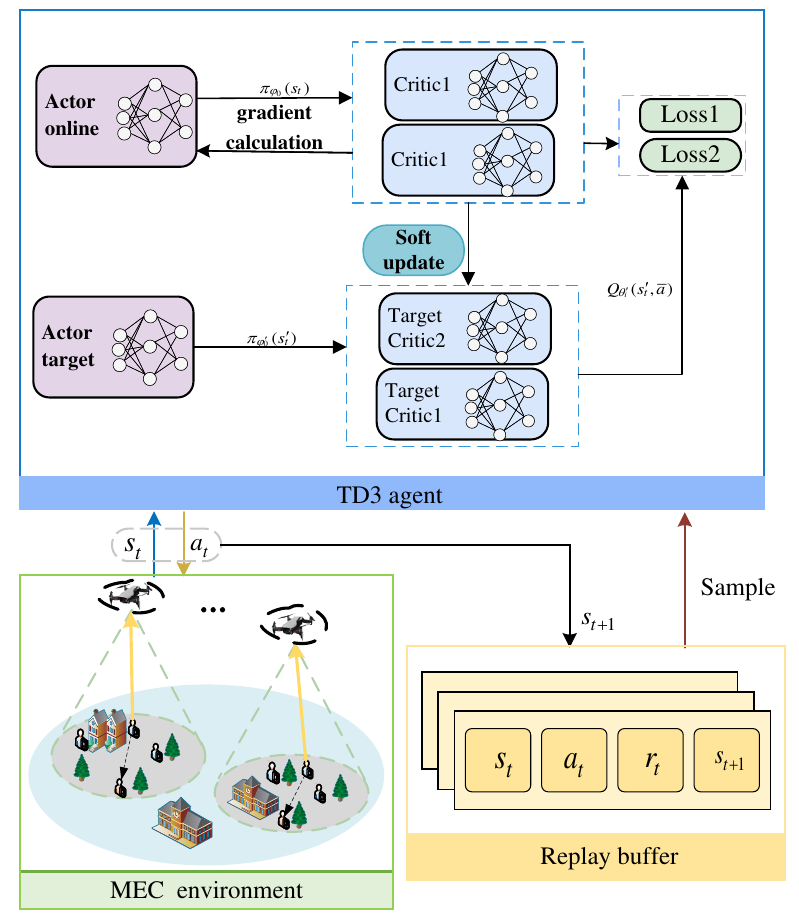}
    \caption{The training framework of TD3.}
    \label{fig:TD3}
\end{figure}
\subsection{TD3-based DRL Training Framework}
In the Actor-Critic framework of DRL, TD3 as a cutting-edge algorithm, includes an actor network and two critic networks, with components denoted by ${\varphi_0}$, ${\theta_1}$, and ${\theta_2}$, respectively. Each network has an associated target network, with components ${\varphi'_0}$, ${\theta'_1}$, and ${\theta'_2}$. During each training iteration, the system randomly samples a minibatch $B_s$ from the experience buffer. For illustration purposes, let $\left( s_t, a_t, r_t, s_{t+1} \right)$ represent a specific transition sample, where $s_{t+1}$ denotes the next state. This section will detail the TD3 method based on Fig. 2.

At each time slot $n$, the agent calculates the output action of the policy network $\pi_{\varphi_0}$ based on the current state $s_n$. By appending clipped Gaussian noise to the target network's output action, the TD3 algorithm presents a regularization strategy, which smooths the output Q-values of the two target critic networks, thereby reducing overfitting. The formula is as follows
\begin{align}
& \bar a = {\pi _{{\varphi _0}}}(s) + clip({\lambda _{\bar a}},{c_{\bar a}}),
\end{align}
where $\text{clip}(\lambda_{\bar{a}}, c_{\bar{a}})$ is a clipping operation that bounds the random variable $\lambda_{\bar{a}}$ within $[-c_{\bar{a}}, c_{\bar{a}}]$, and $\lambda_{\bar{a}} \sim N(0, \varsigma_{\bar{a}}^2)$ represents a zero-mean Gaussian random variable with variance $\varsigma_{\bar{a}}^2$. 
Furthermore, to reduce bias in Q-function estimation, the TD3 algorithm employs a twin Q-function mechanism denoted as $Q_{\theta'_1}$ and $Q_{\theta'_2}$. During gradient descent, the algorithm selects the smaller Q-value between the two outputs, and uses the state-action value function's Bellman expectation equation to compute the TD target
\begin{align}
    & {y_{ta}} = {r_t} + \gamma \min ({Q_{{{\theta '}_1}}}({s_{t+1}},\bar a),{Q_{{{\theta '}_2}}}({s_{t+1}},\bar a)),
\end{align}
where $\gamma$ represents the discount factor. The two target critic networks are optimized by minimizing the loss function $L(\theta_i)$
\begin{align}
    & L({\theta _i}) = \frac{1}{{|B_s|}}{\sum {({y_{ta}} - {Q_{{\theta _i}}}({s_t},{a_t}))} ^2}, i =\{1,2\} .
\end{align}

\begin{algorithm}[t]
	\caption{TD3-based DRL training framework}
	\label{TD3}
	\begin{algorithmic}[1]
		\STATE{Initialize the maximum training episodes $K_{ep}$, training sample length $E$, randomly sampled mini-batch.}
		\STATE{Initialize actor network ${\varphi_0}$, critic network ${\theta_1}$, ${\theta_2}$, target network ${\varphi '_0} \leftarrow {\varphi _0}$, ${\theta '_1} \leftarrow {\theta _1}$, ${\theta '_2} \leftarrow {\theta _2}$, actor learning rate ${l_a}$, critic learning rate ${l_c}$, discount factor $\gamma$, soft update coefficient $\varpi$.}
		\FOR{episode=1 to ($K_{ep}$)}
		\FOR{n =1 to ($E$)}
		\STATE{Obtain action $a_n$ from policy network $\pi_{\varphi}$ based on current state $s_n$, representing allocation ratios for busy UDs, idle UDs, and UAVs. Add Gaussian noise to the action for exploration;}
		\STATE{Execute action, receive reward $r_n$ and next state $s_{n+1}$;}
		\STATE{Store transition $(s_n, a_n, r_n, s_{n+1})$ in experience replay buffer;}
		\STATE{Update current state $s_n$ to $s_{n+1}$;}
		\STATE{Sample mini-batch $B_s$ from experience replay buffer;}
		\FOR{each transition $(s_t, a_t, r_t, s_{t+1})$ in $B_s$}
		\STATE{Compute target action, TD target value, and update Critic network weights.}
		\ENDFOR
		\STATE{Update Actor network parameters ${\varphi_0}$;}
		\STATE{Soft update target network parameters: update $\theta'_i$ using $\theta_i$ and update $\varphi_0'$ using $\varphi_0$.}
		\ENDFOR
		\ENDFOR
	\end{algorithmic}
\end{algorithm}

In the training process of the agent, the primary objective of the agent is to maximize cumulative rewards. Q-values are used to assess the long-term returns of selecting specific actions in the current state. To optimize this process, the actor network needs to generate actions $\pi_{\varphi_0}(s_t)$ that maximize the Q-value. Therefore, the loss function for the actor network is
\begin{align}
    & J({\varphi _0}) = \frac{1}{{|B_s|}}\sum {{Q_{{\theta _1}}}} ({s_t},{\pi _{{\varphi _0}}}).
\end{align}

The critic and actor networks are updated using mini-batch stochastic gradient descent, with the update processes as follows
\begin{align}
    & {\theta _i} \leftarrow {\theta _i} - {l_c}{\nabla _{{\theta _i}}}L({\theta _i}), i =\{1,2\} ,
\end{align}
\begin{align}
    & {\varphi _0} \leftarrow {\varphi _0} - {l_a}{\nabla _{{\varphi _0}}}J({\varphi _0}), i =\{1,2\} ,
\end{align}
where $l_c$ and $l_a$ denote the learning rates for the critic and actor networks, respectively. Finally, the parameters of the actor network $\varphi_0$ and the two critic networks $\theta_1$ and $\theta_2$ are synchronized with their corresponding target network parameters using a soft update method to ensure model stability. The specific target network update strategy is shown below
\begin{align}
    & {\theta '_i} \leftarrow \varpi {\theta _i} + (1 - \varpi ){\theta '_i}, i =\{1,2\} ,
\end{align}
\begin{align}
    & {\varphi '_0} \leftarrow \varpi {\varphi _0} + (1 - \varpi ){\varphi '_0} ,
\end{align}
where $\varpi$ represents the soft update coefficient. Pseudocode is as shown in Algorithm 1.

\subsection{Complexity Analysis}
In this section, we analyze the computing complexity of the proposed TD3 algorithm. The computing complexity of the algorithm is given by 
$\mathcal{O}(\sum\limits_{r = 0}^R {{A_r}{A_{r + 1}}}  + \sum\limits_{s = 0}^S {{C_s}} {C_{s + 1}})$
where \( R \) and \( S \) represent the numbers of fully connected layers in the action and actor networks, respectively. \( A_r \) and \( A_{r+1} \) are the numbers of neurons in the \( r \)-th fully connected layer of the action network and the \( j \)-th fully connected layer of the actor network, respectively. \( C_s \) and \( C_{s+1} \) denote the sizes of input data for the corresponding networks. Therefore, for all \( K_{ep} \) episodes, the time complexity of the training algorithm is 
$O\left( {{K_{ep}}\left( {E\left( {\mathcal{O}(\sum\limits_{r = 0}^R {{A_r}{A_{r + 1}}}  + \sum\limits_{s = 0}^S {{C_s}} {C_{s + 1}})} \right)} \right)} \right)$.

\section{ NUMERICAL RESULTS}
This section demonstrates the effectiveness of the proposed TD3 algorithm in a multi-UAV-assisted D2D MEC network through numerical experiments. TD3 is evaluated against the following benchmark approaches
\begin{itemize}
    \item {\bf{PPO (Proximal Policy Optimization):}} This approach is a widely recognized and reliable DRL algorithm. It iteratively optimizes the policy over multiple time steps to maximize cumulative rewards and introduces a clipping term to limit the extent of policy updates, ensuring the stability of the algorithm.
    \item {\bf{DDPG (Deep Deterministic Policy Gradient):}} This approach leverages the strengths of both deep learning and deterministic policy gradient techniques, making it well-suited for environments with high-dimensional state and action spaces. The DDPG framework consists of a critic target network, an actor target network, a critic network, and an actor network.
\end{itemize}
\subsection{Simulation Settings}
In order to verify the functionality and efficiency of the suggested TD3 training framework in a multi-UAV-assisted D2D MEC network, this section presents simulation findings. The experimental area is a square region of 200$\times 200\, \text{m}^2$, containing 20 randomly distributed UDs and 5 UAVs, with initial positions randomly set within the range of $x, y \in [0, 200]$ meters. With the task size uniformly distributed in the range $[D_{\text{min}}, D_{\text{max}}]$\cite{Dm}. The CPU cycles required by the UDs to process the video task $C[n]$ are within the range $[700, 1500]$. The UAVs can transcode the raw video generated by UDs into five different bitrates, i.e., $c \in \{0.4, 0.8, 1.5, 2.0, 2.3\}$ Mbps. The correlation coefficients for CPU computing cycles are set to $m_1 = 1.54$ and $m_2 = 0.08$, while the CPU energy consumption factors are $s_1 = 10^{-27}$ and $y_1 = 3$. In Table I, additional simulation parameters are provided.
\begin{table}[h]
    \centering
    \captionsetup{labelfont={color=blue}}
    \caption{SIMULATION PARAMETERS }
    \renewcommand{\arraystretch}{1.5} 
    \begin{tabularx}{0.5\textwidth}{XX|XX}
        \hline
        Parameters & Values & Parameters & Values \\
        \hline
        $D_\text{max}$ & 3.5 Mb & $v_\text{max}$ & 25 m/s \\
        $D_\text{min}$ & 1.5 Mb & $\sigma^2$ & -100 dBm \\
        $d_\text{dim}$ & 3 m &  $U_{\rm{tip}}$ & 120 m/s \\
        $H_\text{max}$ & 200 m & $A_c$ & 0.5030 {$\rm{m/s^2}$}\\
        $H_\text{min}$ & 100 m & $v_f$ & 3.6 m/s \\
        $P_a$ & 59.03 W & $p_{d, \text{max}}$& 0.5 W \\
        $P_b$ & 79.07 W  & $f_{k,\rm{max}}$ &30 GHz \\
        $B_0$ & 10 MHZ  & $f_{i,\rm{max}}$ & 1.5 GHz \\
        $B_1$ & 15 MHZ  & $N$ & 50 \\
        $E_r$ & 20 KJ  & $f_{j,\rm{max}}$ & 1.5 GHz \\
        \hline
    \end{tabularx}
\end{table}
\begin{table}[h]
    \centering
    \captionsetup{labelfont={color=blue}}
    \caption{ALGORITHMIC PARAMETERS }
    \renewcommand{\arraystretch}{1.5} 
    \begin{tabularx}{0.5\textwidth}{XX|XX}
        \hline
        Parameters & Values &Parameters & Values\\
        \hline
        $K_{ep}$ & 200 & $E$ & 600 \\
        $\gamma$ & 0.98 & $l_a$ & 0.005 \\
        $\varpi$ & 0.05 & $l_c$ & 0.005 \\            
        \hline
    \end{tabularx}
\end{table}

\subsection{Performance Evaluation}
\begin{figure}[t]
    \centering
    \includegraphics[width=\columnwidth]{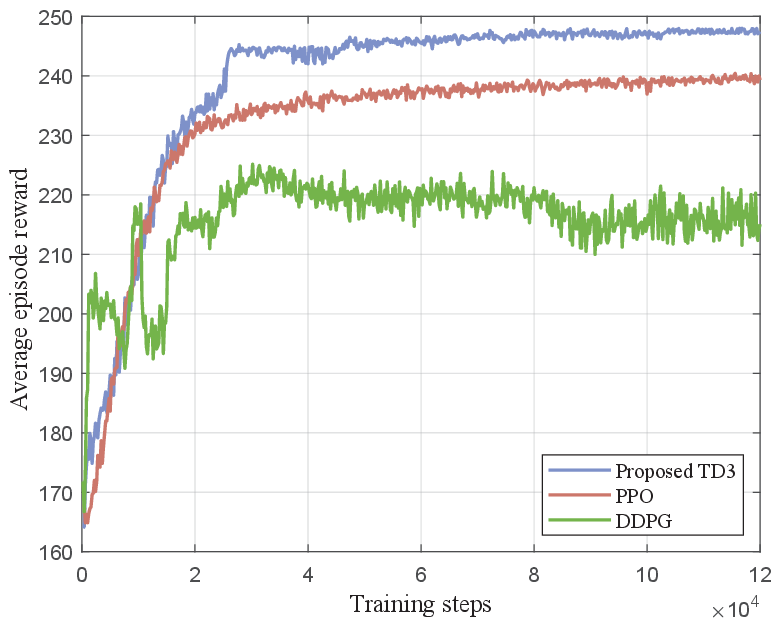}
    \caption{Convergence performance under different algorithms.}
    \label{fig:shoulian}
\end{figure}

In Fig. \ref{fig:shoulian}, the convergence of the TD3 algorithm is compared with other benchmark methods. It is clear that as the quantity of training iterations increases, the reward values of all algorithms gradually improve and tend to converge following a specific amount of steps, demonstrating the validity of DRL methods. It can be observed that after 20,000 steps, the reward value of the TD3 algorithm is significantly higher than that of the other two algorithms, and it converges faster compared to the PPO and DDPG algorithms. This is due to the introduction of a double Q-network and target policy delay update mechanism in TD3, which effectively reduces the bias in Q-value estimation and improves training stability. Additionally, TD3 optimizes exploration and exploitation in continuous action spaces through a noise smoothing strategy, making it more stable in complex environments and tasks. In contrast, DDPG, which uses deterministic action outputs and exploration noise, has weaker exploration capabilities, leading to lower reward values compared to PPO and TD3.

\begin{figure}
    \centering
    \includegraphics[width=\columnwidth]{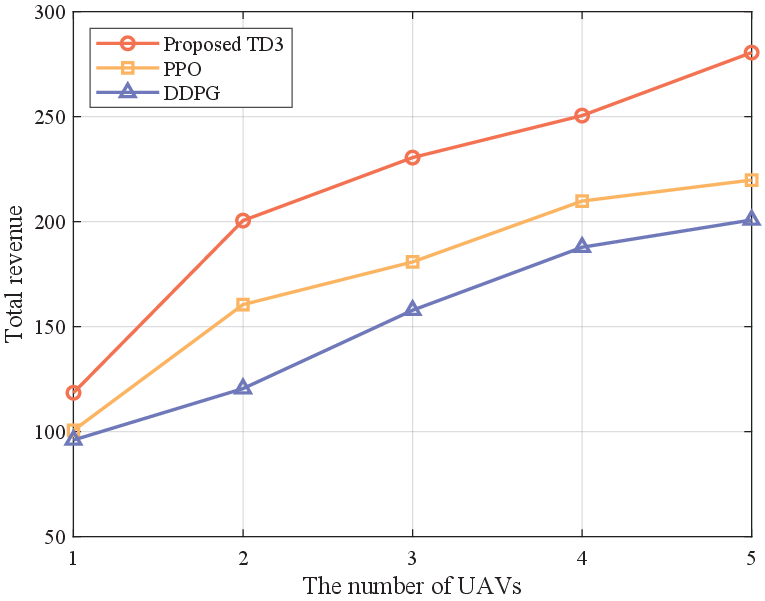}
    \caption{Revenue comparison versus different numbers of UAVs.}  
    \label{fig:UAVtotal}
\end{figure}

Fig. \ref{fig:UAVtotal} compares the rewards of three algorithms with a fixed number of UDs and varying numbers of UAVs. The results show that as the quantity of UAVs increases, the total system reward also increases accordingly. The explanation is that when the quantity of UAVs increases, more busy UDs offload their videos to the UAVs for processing, which boosts the utility revenue of UAVs. The rise in the quantity of UAVs leads to higher costs for busy UDs when purchasing computing resources, and the utility revenue derived from the resources purchased from UAVs also increase. Consequently, the total system reward improves. Additionally, it can be observed that our TD3 approach consistently outperforms both PPO and DDPG algorithms in terms of rewards across all scenarios.

\begin{figure}
    \centering
    \includegraphics[width=\columnwidth]{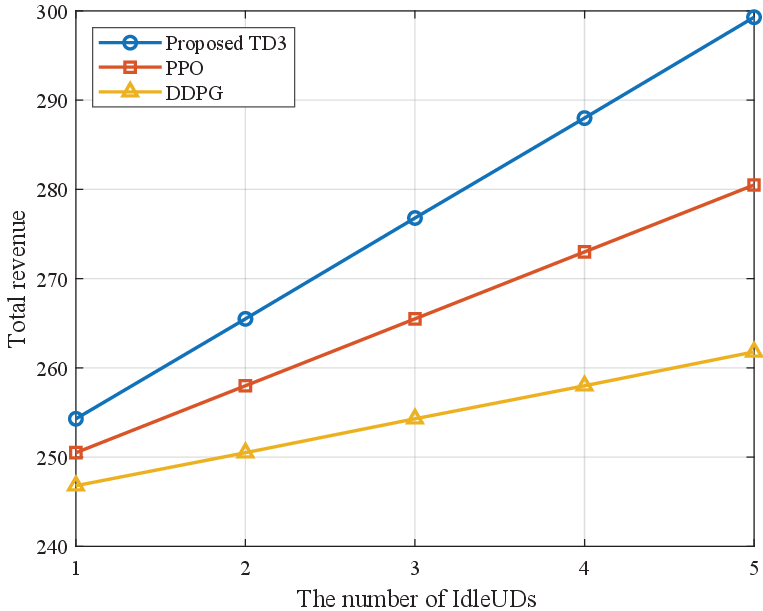}
    \caption{Revenue comparison versus different numbers of IdleUDs.}
    \label{fig:idletotal}
\end{figure}

Fig. \ref{fig:idletotal} compares the rewards of three algorithms with a fixed number of busy UDs and UAVs under varying numbers of idle UDs. The results show that as the amount of idle UDs increases, the total system reward also improves. This is because a rise in the quantity of idle UDs enhances the total reward from collaborative computing. More busy UDs can offload their tasks via D2D, which reduces the energy consumption for local processing by busy UDs, and increases the revenue of demand UDs to some extent, thereby boosting the total reward. Additionally, our TD3 approach outperforms both PPO and DDPG in terms of rewards across all scenarios.

\begin{figure}
    \centering
    \includegraphics[width=\columnwidth]{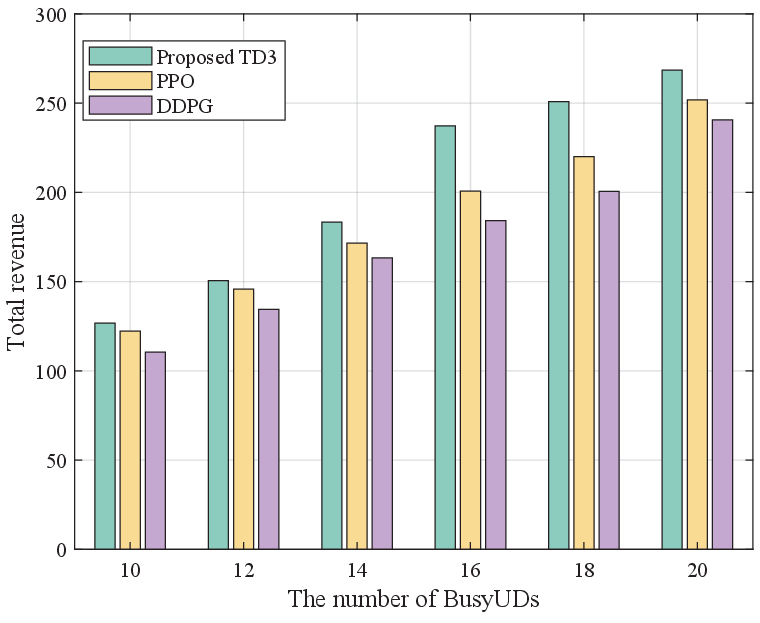}
    \caption{Revenue comparison versus different numbers of BusyUDs.}
    \label{fig:busytotal}
\end{figure}

Fig. \ref{fig:busytotal} evaluates the rewards of three algorithms under various amounts of busy UDs, while keeping the number of idle UDs and UAVs fixed. The results show that when the amount of busy UDs rises, the total system reward also increases. This is because more busy UDs need to process video tasks and offload them to UAVs or idle UDs for cooperative computing, which enhances the overall system utility. As the number of busy UDs increases, the utilization of computing resources provided by UAVs and idle UDs improves, further promoting an increase in total system reward. 
\begin{figure}
    \centering
    \includegraphics[width=\columnwidth]{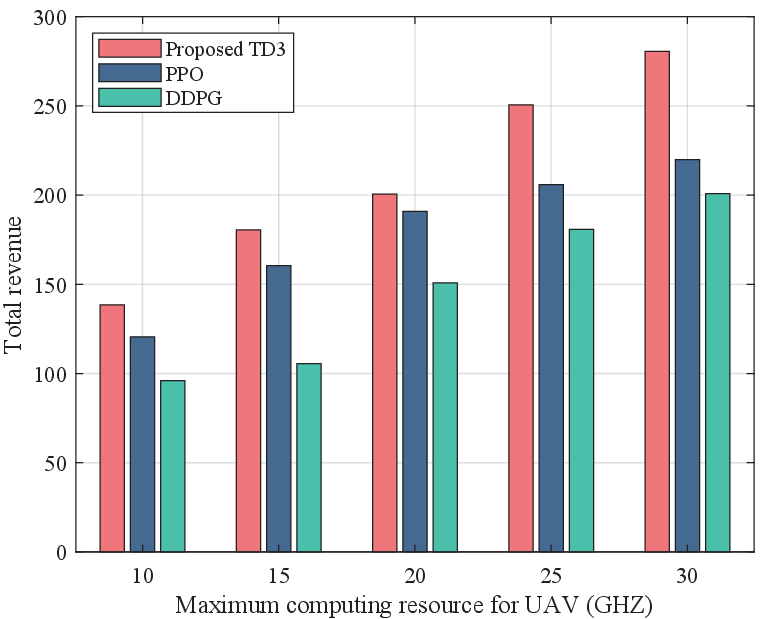}
    \caption{Impact of maximum computing resources for UAVs on total revenue.}
    \label{fig:comput}
\end{figure}

Fig. \ref{fig:comput} shows the effect of the maximum computing resources of the UAV on the total revenue. As the maximum computing resource of the UAV increases, the total revenue increases. This is because as the total maximum computing resources of the UAV increase, the available computing resources of the UAV also increase, BusyUD $i$ can offload more locally generated video tasks to the UAV for processing, the more BusyUD $i$ get from purchasing computing resources, the cost of local processing is saved, and the total system revenue increases.

\begin{figure}
    \centering
    \includegraphics[width=\columnwidth]{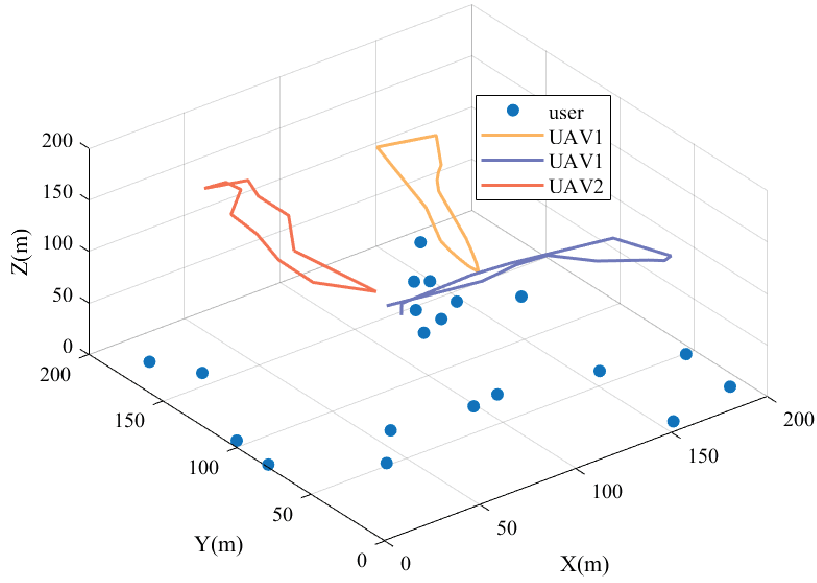}
    \caption{The UAVs' trajectories example with $I=20$ and $K=3$.}
    \label{fig:tr}
\end{figure}
In Fig. \ref{fig:tr}, we present the trajectories of UAVs for $I = 20$ and $K = 3$.
The figure demonstrates that each UAV adjusts its flight path to approach user clusters, thereby effectively reducing the distance to UDs. This strategy not only optimizes the energy consumption for communication and computation offloading but also enhances the overall system revenue.
 
\section{CONCLUSION}
This paper proposed a multi-UAV-assisted D2D MEC framework for video streaming, aiming to maximize system revenue by optimizing UAV power allocation, video transcoding strategy, computing resource pricing, task offloading ratios, and the flying trajectory. To deal with this challenging optimization problem, it was modeled as an MDP and solved by capitalizing on the TD3 algorithm. Simulation results indicated that the TD3 algorithm performs exceptionally well in improving system utility, significantly outperforming other comparative methods, thus validating its effectiveness and advantages. Future research will further explore how  to optimize task offloading ratios and resource pricing through multi-agent algorithms in complex scenarios with multiple UAVs and multiple base stations, in order to further enhance the overall system benefits.

\bibliographystyle{IEEEtran}
\bibliography{IEEEabrv,refs}

\end{document}